# Reentrant quantum anomalous Hall effect in molecular beam epitaxy-grown MnBi$_2$Te$_4$ thin films

Yuanzhao Li[1,2]†, Yunhe Bai[2]†, Yang Feng[1]†*, Jianli Luan[1,2]†, Zongwei Gao[1], Yang Chen[2], Yitian Tong[2], Ruixuan Liu[2], Su Kong Chong[1,3], Kang L. Wang[3,4], Xiaodong Zhou[5,6,7], Jian Shen[5,6,7,8,9], Jinsong Zhang[2,14,15], Yayu Wang[2,14,15], Chui-Zhen Chen[10,11], XinCheng Xie[6,12,15], Xiao Feng[1,2,14,15]*, Ke He[1,2,14,15]*, and Qi-Kun Xue[1,2,13,14,15]

[1]Beijing Academy of Quantum Information Sciences, Beijing 100193, China.
[2]State Key Laboratory of Low Dimensional Quantum Physics and Department of Physics, Tsinghua University, Beijing 100084, China.
[3]Department of Electrical and Computer Engineering, University of California Los Angeles, Los Angeles, California 90095, USA.
[4]Department of Physics and Astronomy, University of California Los Angeles, Los Angeles, California 90095, USA.
[5]State Key Laboratory of Surface Physics and Department of Physics, Fudan University, Shanghai 200438, China.
[6]Institute for Nanoelectronic Devices and Quantum Computing, Fudan University, Shanghai 200438, China.
[7]Zhangjiang Fudan International Innovation Center, Fudan University, Shanghai 201210, China.
[8]Shanghai Research Center for Quantum Sciences, Shanghai 201315, China.
[9]Collaborative Innovation Center of Advanced Microstructures, Nanjing 210093, China.
[10]School of Physical Science and Technology, Soochow University, Suzhou 215006, China.
[11]Institute for Advanced Study, Soochow University, Suzhou 215006, China.
[12]International Center for Quantum Materials, School of Physics, Peking University, Beijing 100871, China.
[13]Southern University of Science and Technology, Shenzhen 518055, China.
[14]Frontier Science Center for Quantum Information, Beijing 100084, China.
[15]Hefei National Laboratory, Hefei 230088, P. R. China.

*Corresponding author. Email : fengyang@baqis.ac.cn ; xiaofeng@mail.tsinghua.edu.cn ; kehe@tsinghua.edu.cn
†These authors contributed equally to this work.

## Abstract

In the realm of topological quantum physics, it is widely accepted that a global bandgap is necessary to achieve quantized Hall conductance, irrespective of its origin—magnetic field, exchange coupling or other mechanisms. A material exhibiting the quantum anomalous Hall (QAH) effect had been considered equivalent to a Chern insulator, a band insulator with non-zero Chern number. However, a **Chern Anderson insulator** (CAI), which arises in a conducting state with exchange-induced Berry curvature splitting and disorder-induced Anderson localization, can in principle exhibit the QAH effect. In this study, we investigate intrinsic magnetic topological insulator MnBi$_2$Te$_4$ thin films grown by molecular beam epitaxy (MBE). We observe a reentrant QAH effect when the Fermi energy enters the valance band and magnetic field equals zero, indicating the emergence of the CAI state. The discovery opens a new avenue for realizing the QAH effect and underscores the fundamental role of both Berry curvature and Anderson localization.

# Main text

The quantum anomalous Hall (QAH) effect displays quantized Hall conductance and vanishing longitudinal one in the absence of external magnetic field [1]. After its first realization in a magnetic topological insulator (MTI) Cr-doped $(Bi,Sb)_2Te_3$ [2], intensive efforts have been made to improve the sample quality by reducing the inevitable randomness of magnetic dopants. The QAH effect has been reproduced in various systems, such as MTI with modulation magnetic doping [3], MTI with various magnetic dopants [4,5], intrinsic MTI [6], two-dimensional (2D) moiré superlattice without magnetic elements [7–10], and spin-orbit proximitized graphene [11,12]. More recently, the fractional counterpart of QAH effect has been observed in twisted transition metal dichalcogenide layers [13,14] and high-quality graphene-based moiré superlattices [15], which offers an unprecedented opportunity to investigate strongly correlated topological electronic states. The materials with topologically protected quantum state, combined with properties such as the inherent spin-momentum locking, may lead to new device concepts for electronic applications.

The $MnBi_2Te_4$ [16–20] is anticipated to significantly enhance the magnetism with its high concentration and ordered magnetic atoms. The material has an A-type antiferromagnetic (AFM) order, resulting from the coupling of intralayer ferromagnetic (FM) and interlayer AFM. Consequently, thin odd-layer $MnBi_2Te_4$ possesses an uncompensated magnetization, and is expected to exhibit QAH effect at high temperatures, owning to its topologically nontrivial band structure with a large exchange gap [17,18]. However, the zero-magnetic-field (ZMF) QAH effect is seldom observed in few-odd-layer $MnBi_2Te_4$, and the predicted large gap for the QAH state (~50 meV) has not been identified in spectroscopy [21–25] . The reason has been attributed to complex magnetic phases [26], multidomain structure [23,27], anti-site defects [28–31], or nonuniform thickness [32], but no consensus is concluded. The situation has recently led to speculation that the ZMF-state of five-layer $MnBi_2Te_4$ in most samples, with large longitudinal resistance and small Hall signal, might be a topologically trivial insulator [33] due to magnetic disorder. The robustness of the QAH effect and its relation to the Berry curvature, magnetism, and electronic ground states in the ZMF-state of $MnBi_2Te_4$ remain largely mysterious.

In this work, we investigate molecular beam epitaxy (MBE)-grown $MnBi_2Te_4$ thin films which show the QAH state at high magnetic field and become topologically trivial at zero magnetic field. Surprisingly, the films reenter the QAH state when the Fermi level is tuned into the valence band. More importantly, in contrast to the conventional wisdom that a QAH system exhibiting nonzero integer Hall conductance must be a Chern insulator with global bandgap [1,32], the observed reentrant QAH effect occurs in a gapless system subject to Anderson localization [34,35] which can thus be called a Chern Anderson insulator (CAI).

As a matter of fact, the stability of topological states against local perturbations is protected by an energy gap, irrespective of whether the bandgap is induced by external magnetic field [36,37], exchange coupling with local magnetic moments [38,39], or complex interplay between correlation effect and hopping interactions [40–44]. Otherwise the system under adiabatic evolution [45] will undergo a topological phase transition when the gap between different eigenstates closes. Specifically, in the thin films of a time-reversal symmetry (TRS) protected three-dimensional topological insulator [46], the gapless top and bottom surface states are weakly hybridized through tunneling effect and double-degenerated [38]. The formation of long-range magnetic order brings the exchange coupling between electron spins and local moments of magnetic ions, which leads to the exchange-induced band splitting between different components of spin texture and gap opening near the Dirac point (DP) [38]. If the initial hybridized surface bands are already inverted, a sufficiently large average exchange energy ($E_{ex}$) will release the inversion between the inner pair of bands, but enlarge the band inversion between the outer pair (Fig. 1A). In this situation, the Berry curvature carried by the inner pair of bands ($\Omega_{in}$) almost vanishes completely, while the other ($\Omega_{out}$) has nonzero values in its spectrum (Fig. 1B). Correspondingly, the inner-bands-carried Chern number ($C_{in}$) defined as integral of $\Omega_{in}$ [47] keeps zero, but the outer-bands-carried Chern number ($C_{out}$) has nontrivial value within the final outer gap (Fig. 1C). This suggests the formation of a Chern insulator state with the presence of the QAH effect [38] when the Fermi energy ($E_F$) lies inside the global bandgap (grey shaded area, $V_0$).

Such a Chern insulator state is observed under the high-magnetic-field (HMF) polarized FM order in our MnBi$_2$Te$_4$ thin films, which are grown on sapphire substrate with *in-situ* oxygen exposure and post-annealing process as described in our previous work [48]. Figure 1D displays gate voltage ($V_g$) dependence of Hall resistivity ($\rho_{yx}$) and longitudinal resistivity ($\rho_{xx}$) of the five-layer thin film #1 performed at -9 T and 20 mK. Within the $V_g$ range from -6 V to -3.5 V as denoted by the grey shaded area, the wide $\rho_{yx}$ plateau with a mean of 1.0002 $h/e^2$ and the vanishing $\rho_{xx}$ with a mean of 0.0018 $h/e^2$ characterize an HMF-QAH effect, which is consistent with previous reports in thin flakes [6,49,50,33]. Hereinafter we assign -5 V as the charge neutral point (CNP) of HMF-QAH regime, which corresponds to $V_0$ inside the global bandgap in Fig. 1C. Notably, $\rho_{xx}$ = 0.0018 $h/e^2$ (46.2 Ω) at $V_0$ is dramatically reduced from 5.16 kΩ previously reported in MnBi$_2$Te$_4$ thin films [48], indicating a significant improvement of the sample quality. In the range from -10.5 V to -8.5 V denoted by the gold shaded area away from $V_0$, the simultaneously increasing $\rho_{xx}$ and reducing $\rho_{yx}$ indicate breakdown of HMF-QAH effect when $E_F$ leaves the global bandgap.

However, the sufficiently large $E_{ex}$ is not naturally fulfilled under the A-type AFM order of MnBi$_2$Te$_4$ [47] due to spatial nonuniformity of magnetization and reduction of $E_{ex}$ by anti-site defects [28–31]. As suggested by a practical model for MTI [35], a nuanced yet significant scenario emerges when the average $E_{ex}$ is weak relative to magnetic disorder. Figure 1E illustrates the band structure, where both inner and outer pairs of hybridized bands remain inverted due to weak exchange-induced splitting and the final gaps are topologically nontrivial. The contribution to both $\Omega_{in}$ and $\Omega_{out}$ primarily originates from the states near Γ point with small momentum in **k**-space, and are likely to be confined within a narrow energy range [47]. This results in a peak in the Berry curvature spectrum near the band edge (Fig. 1F). Given the electron-hole asymmetry that the energy dispersion of surface states in the (Bi, Sb)$_2$Te$_3$ materials below the DP is smaller than that above the DP, the separation of $\Omega_{in}$ and $\Omega_{out}$ peaks due to band splitting below the DP in valence band is more pronounced. Simultaneously, the total Chern number $C = C_{in} + C_{out}$ no longer equals nontrivial integer at $V_0$ (Fig.

1G). Instead, $C = 1$ when $E_F$ enters the valence band, as indicated by the gold shaded area ($V_\Omega$). Ultimately, the exchange-induced Berry curvature splitting and disorder-induced Anderson localization can, in principle, give rise to a Chern Anderson insulator (CAI) state without a global bandgap. This state is marked by r-QAH effect, which refers to a QAH effect at 0 T and $V_\Omega$ [35] that is experimentally observed in this work.

To understand the observed ZMF-QAH effect, we investigate its $V_g$-dependent $\rho_{yx}$ and $\rho_{xx}$ at 20 mK in Fig. 1H. A maximum of $\rho_{xx} \approx 4.4$ $h/e^2$ appears at -4.7 V, indicating that the range marked by the gray shaded area around -5 V indeed corresponds to the global bandgap where conduction is mostly prohibited. The suppression of conduction is also reflected in the noisy $\rho_{yx}$ in the same gate range. Intriguingly, the overall profile of ZMF $\rho_{yx}$ with respect to $V_g$ is very similar to the energy spectrum of $C$ in Fig. 1G. Within the range from -10.5 V to -8.5 V denoted by the gold shaded area, $\rho_{yx}$ arises and ultimately exhibits a nearly quantized plateau, indicating a ZMF-QAH effect when $E_F$ enters the valence band. In other words, the r-QAH effect is obtained in the MnBi$_2$Te$_4$ thin film. We assign -9.5 V to the center of the r-QAH regime, corresponding to $V_\Omega$ in Fig. 1G. This r-QAH effect is further confirmed by the simultaneous decrease of $\rho_{xx}$ from the maximum near $V_0$ to a dip ($\rho_{xx} \approx 0.85$ $h/e^2$) near $V_\Omega$. Furthermore, the occurrence of r-QAH effect at 0 T and $V_\Omega$ is also consistent with the electron-hole asymmetry in (Bi, Sb)$_2$Te$_3$ materials as emphasized in Fig. 1E. An immediate interpretation of these results is that the QAH effect can emerge from conducting states, signifying the transition from an ZMF-insulating state at $V_0$ to a CAI state at $V_\Omega$.

Shown in Figure 2 is the magnetic field ($\mu_0 H$) dependence of $\rho_{yx}$ and $\rho_{xx}$ at $V_0$ and $V_\Omega$. For the $V_0$ case, the HMF-QAH effect is further confirmed in range 6 T < |$\mu_0 H$| < 9 T, seen by fully quantized $\rho_{yx}$ with a mean of 1.0004 $h/e^2$ in Fig. 2A and a vanishing $\rho_{xx}$ with a mean of 41.0 Ω ($\approx 0.0016$ $h/e^2$) in Fig. 2B. As $\mu_0 H$ decreases, $\rho_{yx}$ deviates from the quantized plateau, whereas $\rho_{xx}$ increases monotonously, illustrating a transition from HMF-FM-QAH state to a ZMF-AFM-insulating one. The latter has two typical features: one is the hysteretic $\rho_{yx}$ and $\rho_{xx}$ with coercivity $\mu_0 H_c = 0.90$ T due to the finite net magnetization corresponding to the odd-layer number; the other is a ZMF |$\rho_{yx}$| $\approx 0.6$ $h/e^2$ and a large $\rho_{xx} \approx 7.6$ $h/e^2$, indicating the breakdown of QAH effect.

For the $V_\Omega$ case, the $\mu_0 H$-dependent $\rho_{yx}$ and $\rho_{xx}$ in Fig. 2C and Fig. 2D undoubtedly confirms the reentrance picture. The entire demagnetization process can be divided into several stages. The first stage is the reinforcing process from ±9 T to ±7.2 T (orange ticks), in which |$\rho_{yx}$| persists full quantization while $\rho_{xx}$ drops from $\rho_{xx} \approx 0.07$ $h/e^2$ to $\rho_{xx} \approx 0.04$ $h/e^2$. The second stage is the rapid-weakening process from ±7.2 T (orange ticks) to ±3.8 T (red ticks), in which the perfect quantization drops off to |$\rho_{yx}$| $\approx 0.9$ $h/e^2$ and $\rho_{xx}$ rapidly increases up to $\rho_{xx} \approx 1.0$ $h/e^2$. The third stage is the gradual-weakening process from ±3.8 T (red ticks) to ±2.1 T (green ticks), in which |$\rho_{yx}$| is further reduced to |$\rho_{yx}$| $\approx 0.8$ $h/e^2$ and $\rho_{xx}$ keeps increasing to $\rho_{xx} \approx 1.3$ $h/e^2$. The last stage is the reentering process from ±2.1 T (green ticks) to 0 T, at the end of which |$\rho_{yx}$| regains a nearly quantized value 0.9736 $h/e^2$, which represents the most important finding of this work. Note that, the ZMF $\rho_{xx} = 1.0756$ $h/e^2$ in Fig. 2D is larger than $\rho_{xx} \approx 0.85$ $h/e^2$ at $V_\Omega$ in Fig. 1H, which is likely caused by the heating effect during the $\mu_0 H$ sweep process. For comparison, ZMF $\rho_{yx}$ and $\rho_{xx}$ of the QAH effect in the exfoliated MnBi$_2$Te$_4$ thin flake is 0.97 $h/e^2$ and 0.061 $h/e^2$, respectively [6].

To further explore the r-QAH effect, we systematically compare the transport behavior of the ZMF-insulating state (0 T and $V_0$) and the CAI state (0 T and $V_\Omega$). Fig 3A (3B) show the temperature ($T$) dependence of ZMF $\rho_{yx}$ and $\rho_{xx}$ at $V_0$ ($V_\Omega$). In both cases, $\rho_{yx}$ exhibits larger value at lower $T$, but the trend of $\rho_{xx}$ is different. The ZMF-insulating state displays a diverging $\rho_{xx}$ as $T$ approaches zero. In contrast, the CAI state shows a metallic behavior below 0.5 K, as characterized by the reducing $\rho_{xx}$ as $T$ decreases. This provides additional evidence for the breakdown (reentrance) of QAH effect at $V_0$ ($V_\Omega$).

Figure 3C and 3D present an instructive analysis by plotting the $\mu_0 H$-dependent tangent of the Hall angle $\theta_H$, defined as $\tan\theta_H = \sigma_{xy}/\sigma_{xx} = \rho_{yx}/\rho_{xx}$ [51], where $\sigma_{xy}$ ($\sigma_{xx}$) denotes Hall (longitudinal) conductivity. For $V_0$ case (Fig. 3C), $\tan\theta_H$ decays monotonously from a large value exceeding 400 (indicative of HMF-QAH effect) to a small value below 0.1, suggesting a ZMF-insulating state. This decayed Hall response with reducing $\mu_0 H$ is consistent with the $\mu_0 H$-dependent $\rho_{yx}$ at $V_0$ in Fig. 2A. In contrast, $\tan\theta_H$ at $V_\Omega$ (Fig. 3D) exhibits a non-monotonic behavior, in which four stages (reinforcing, rapid-weakening, gradual-weakening, and reentering) are clearly resolved. This multistage Hall response at $V_\Omega$ confirms the evolution from the HMF-QAH effect to the r-QAH effect, providing valuable insights into the formation of the CAI state.

To elucidate the energetics of the transport behavior, we employ line fits to the Arrhenius plots of $1/T$-dependent $\sigma_{xx}$ [47]. This yields the energy scale $\Delta E$ at $V_0$ (Fig. 3E) and $V_\Omega$ (Fig. 3F). It characterizes the minimum energy required to trigger the thermal excitation of electrons participating in global transport. When $E_F$ lies inside the global bandgap, $\Delta E$ equals the spacing of nearest band edges. However, when $E_F$ enters the valence band with the presence of disorder, $\Delta E$ describes the mobility gap. It is important to note that in the QAH regime, $\sigma_{xx} = \rho_{xx}/(\rho_{xx}^2 + \rho_{yx}^2)$ no longer equals $1/\rho_{xx}$, hence $\Delta E$ cannot reflect the metallic behavior ($\rho_{xx}$ vs $T$) in Fig. 3B directly. The common feature in Fig. 3E and Fig. 3F is the minimum $\Delta E$ occurring near $|\mu_0 H| = 3.8$ T (red tick), corresponding to the transition from HMF-FM order to ZMF-AFM order. On both sides of the red tick, $\Delta E$ shows monotonic changes at $V_0$, but displays non-monotonic $\mu_0 H$ evolution at $V_\Omega$. Two important differences are particularly noted. First, an inflection point or maximum of $\Delta E$, which is absent in $V_0$ case, appears at $|\mu_0 H| = 7.2$ T (orange tick) in $V_\Omega$ case, closely resembling $|\mu_0 H| \approx 7.6$ T in the exfoliated flake with ZMF-QAH effect [6]. Second, within the reentrant regime $|\mu_0 H| < 2.1$ T, instead of increasing at $V_0$, $\Delta E$ decreases as $\mu_0 H$ approaches zero at $V_\Omega$. Overall, the $\mu_0 H$ evolution with four stages (marked by ticks) are consistent with $\mu_0 H$-dependent $\rho_{yx}$, $\rho_{xx}$, $\tan\theta_H$, and $\Delta E$ (Fig. 2 and Fig. 3), which shows contrasting behavior between $V_0$ and $V_\Omega$ case.

In Fig. 3G, we give a phase diagram in the intensity plot of $\rho_{yx}$ as a function of $\mu_0 H$ and $V_g$, offering a comprehensive view into the evolution of the QAH state in our system. The direction of $\mu_0 H$ sweep is from -9 T to +9 T, as indicated by the black arrow. The HMF-QAH regimes primarily exist at $|\mu_0 H| >$ 4 T and $V_g < -2$ V. In contrast, the r-QAH regime, characterized by a strong rebuilding of $\rho_{yx} \approx 0.97$ $h/e^2$ only exists in a narrow range around 0 T and $V_\Omega$ (gold tick), consistent with the electron-hole asymmetry (Fig. 1E and 1F). The presence of the r-QAH effect, as well as the ZFM-insulating state, is further revealed by the intensity plot of $\rho_{xx}$ [47].

A more intuitive understanding of the r-QAH effect at 0 T and $V_\Omega$ can be gleaned by comparing film #1 with other ZMF results. In particular, we include three additional MBE-grown thin films #2, #3, #4 [47], a seven-layer exfoliated MnBi$_2$Te$_4$ thin flake #5 with pressure-enhanced ZMF-QAH effect [52], and a five-layer thin flake #6 exhibiting ZMF-QAH effect [6]. Fig. 4A displays the $T$-dependent tan$\theta_H$ for all samples. Note that, except for #1, all the data in this plot is at each CNP ($V_0$). A contrasting behavior is observed between films (#2, #3, #4) and flakes (#5, #6). Specifically, as $T$ approaches zero, the tan$\theta_H$ curve of #2, #3, #4 decreases whereas those of #5, #6 diverge. Interestingly, the behavior of film #1 at $V_0$ aligns more closely with sample #2, #3, #4, whereas at $V_\Omega$ it shares similar upward trend with #5, #6. It strongly suggests that, the MnBi$_2$Te$_4$ thin film #1 undergoes a topological transition from a ZMF-insulating state to a CAI state exhibiting r-QAH effect with the aid of $V_g$.

We further undertake a scaling analysis of ZMF tan$\theta_H$ with respect to $T$ and $\Delta E$. As depicted in Fig. 4B, all tan$\theta_H$ versus $T/\Delta E^\alpha$ curves, both within the ZMF-insulating regime (#1@$V_0$, #2, #3, #4) and the QAH regime (#5, #6), coalesce into single curves with $\alpha \approx 0.6 \pm 0.05$. This excludes the extremely low $T$ regime where lattice temperature and electron temperature typically decouple. The transition point can be obtained at tan$\theta_H \approx 0.5$ where a $T$-independence is observed [53]. However, the tan$\theta_H$ curves in the QAH regime (#5, #6) and the r-QAH regime (#1@$V_\Omega$) are not directly overlapped, leaving an open question of whether the r-QAH effect in the CAI state and the QAH effect in Chern insulator state [6,52] are entirely equivalent. One possible difference might be concealed in the fact that a bandgap and a mobility gap are not directly comparable. Nevertheless, future theoretical and experimental endeavors are needed to scrutinize such a CAI state and develop a proper theory describing its scaling behavior.

Figure. 4C provides another perspective to understand the CAI state by viewing the evolution in the parameter plot of conductivities ($\sigma_{xy}(\beta)$, $\sigma_{xx}(\beta)$), where $\beta$ involves two parameters, $T$ and $V_g$. We first analyze the ZMF behavior of samples prior to this work (#2 to #6) by examining the $T$-dependent conductivity ($\sigma$) tensor (dashed lines). Under the perspective of renormalization group (RG) flow, the $\sigma$ tensor under broken TRS is suggested to flow towards critical points in the condition of diverging system size ($L$). Such condition is experimentally equivalent to cooling the system ($T \rightarrow 0$) where the inelastic scattering length or phase coherence length diverges. As $T$ approaches zero, thin films #2, #3, #4 flow to (0, 0) consistent with a ZMF-insulating state. Conversely, exfoliated thin flakes #5, #6 flow towards ($e^2/h$, 0) as anticipated for a QAH state. These two resolved stable phases are consistent with previous work [54–56].

We further analyze the evolution of film #1 from $V_0$ to $V_\Omega$ denoted by solid line in Fig. 4C, which is adapted from Fig. 1H. It collapses onto a semicircle of radius $e^2/2h$ centered at ($e^2/2h$, 0), consistent with the open circles extracted from $\mu_0 H$ sweeps in Fig. 3G. This recalls the semicircular law, which is expected as long as $\rho_{yx}$ is quantized irrespective of $\rho_{xx}$ [57,58], and has been utilized to describe the transition from the insulating state to a QAH state [56]. Clearly, in such a $V_g$-driven transition, $\sigma$ tensor flows from a ZMF-insulating state towards a QAH state and ends on the right side of ($e^2/2h$, $e^2/2h$). Notably, the transition point (indicated by blue arrow) with $T$-independent $\tan\theta_H \approx 0.5$ obtained from Fig. 4B is not in the vicinity of ($e^2/2h$, $e^2/2h$), as predicted by a modular symmetry group theory [59]. Similar asymmetric RG flow has been reported in MTI system [60] and even in an ideal quantum well sample with perfect semicircle relation [58]. The two distinct ground states of film #1 on both sides of transition point are also confirmed by their $T$-dependent $\sigma$ flows at $V_0$ and $V_\Omega$ denoted by green and blue dashed lines in Fig. 4C.

In summary, we have observed the r-QAH effect in MBE-grown $MnBi_2Te_4$ thin films, which signifies the formation of the CAI state. Different from previously reported QAH states that relies on a global bandgap, the newly found QAH state exists in a conducting state with a moderate exchange level and weak disorder. The exchange-induced Berry curvature splitting and Anderson localization both contribute to this significant phenomenon. Our work not only paves a new path for realizing QAH effect, but also stimulates research to explore Berry curvature-related physical phenomena in multiband systems without a global bandgap. This includes chiral optical responses, topological plasmonic and excitonic polaritons in metals, semiconductors, and superconductors with unconventional pairing.

## Acknowledgement

We thank Junjie Qi, Peng Deng, and Chang Liu for helpful discussions.

# Figures

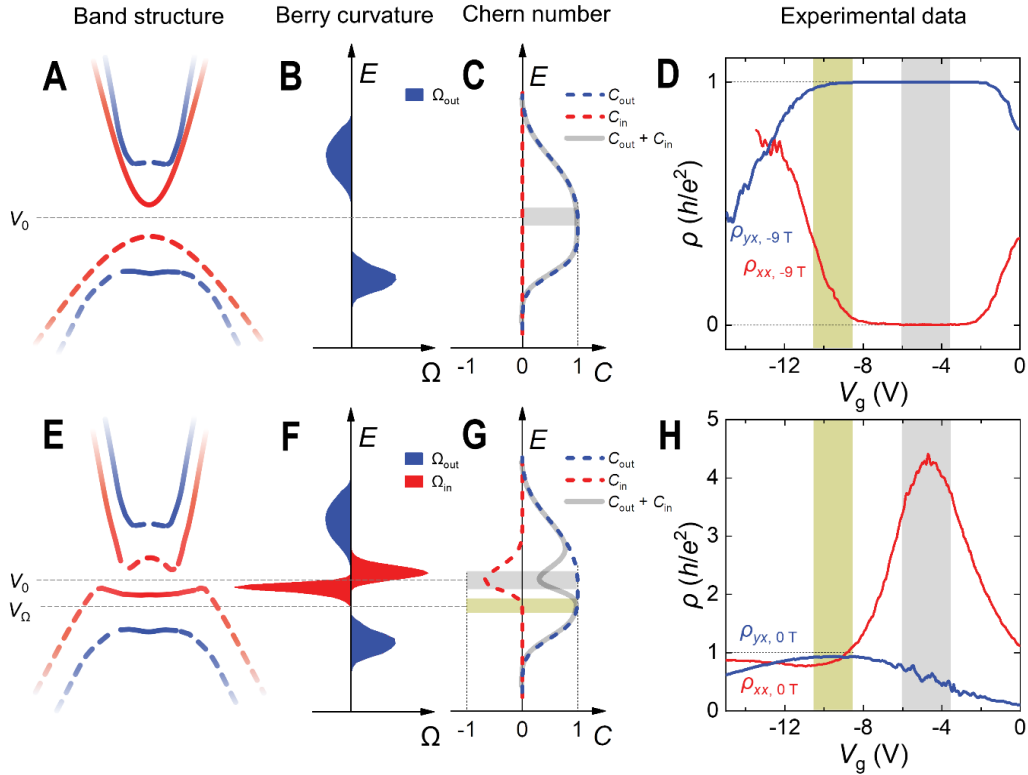

**Fig. 1. Scheme of band structure, Berry curvature, Chern number and experimental performance of an MTI upon reducing the $E_{ex}$.** The initial bands without $E_{ex}$ are topologically nontrivial and double degenerated. Red or blue color denotes different component of spin texture. The solid or dashed line denotes the opposite parity at $\Gamma$ point. $E_{ex}$ draws the inner pair of bands (red) closer and pushes the outer pair (blue) farther away. The different energy dispersions of conduction and valence bands are due to the electron-hole asymmetry in MTI. (**A**) Band structure with strong $E_{ex}$. A large enough $E_{ex}$ releases the nontrivial gap between the inner pair of bands. (**B**) The spin-resolved Berry curvature carried by inner and outer bands ($\Omega_{in}$ and $\Omega_{out}$) with strong $E_{ex}$. $\Omega_{in}$ vanishes, but $\Omega_{out}$ is nonzero with two opposite peaks. (**C**) The spin-resolved Chern number ($C_{in}$ and $C_{out}$) with strong $E_{ex}$. $C_{in}$ vanishes but $C_{out}$ is nonzero. The grey shaded area ($V_0$) denotes the CNP inside the global bandgap, which exhibits a Chern insulator state marked by a QAH effect $\sigma_{xy} = (C_{in} + C_{out}) e^2/h = e^2/h$. (**D**) $V_g$-dependent $\rho_{yx}$ and $\rho_{xx}$ of film #1 acquired at $\mu_0 H$ = -9 T. The grey shaded area denotes HMF-QAH regime (-6 V, -3.5 V). $V_0$ = -5 V is assigned to the CNP of HMF-QAH regime. (**E**) Band structure with weak $E_{ex}$. Both pairs of bands remain topologically nontrivial, with the presence of magnetic disorder and insufficient $E_{ex}$. (**F**) Spin-resolved Berry curvature with weak $E_{ex}$. Both $\Omega_{in}$ and $\Omega_{out}$ have two opposite peaks. (**G**) Spin-resolved Chern number with weak $E_{ex}$. Both $C_{in}$ and $C_{out}$ are nonzero. At $V_0$ inside the gap, $C_{in} + C_{out} \neq 1$. The gold shaded area denotes the CAI regime ($V_\Omega$), which refers to a conducting state with exchange-induced Berry curvature splitting and disorder-induced localization, marked by a r-QAH effect $C_{in} + C_{out} = 1$. (**H**) $V_g$-dependent $\rho_{yx}$ and $\rho_{xx}$ of film #1 acquired at $\mu_0 H$ = 0 T. Within (-6 V, -3.5 V), the ZMF $\rho_{yx}$ is noisy, and the maximum $\rho_{xx} \approx 4.4 \, h/e^2$ appears at -4.7 V, suggesting a ZMF-insulating state. The gold shaded area denotes the r-QAH regime (-10.5 V, -8.5 V). $V_\Omega$ = -9.5 V is assigned to the center of the r-QAH regime.

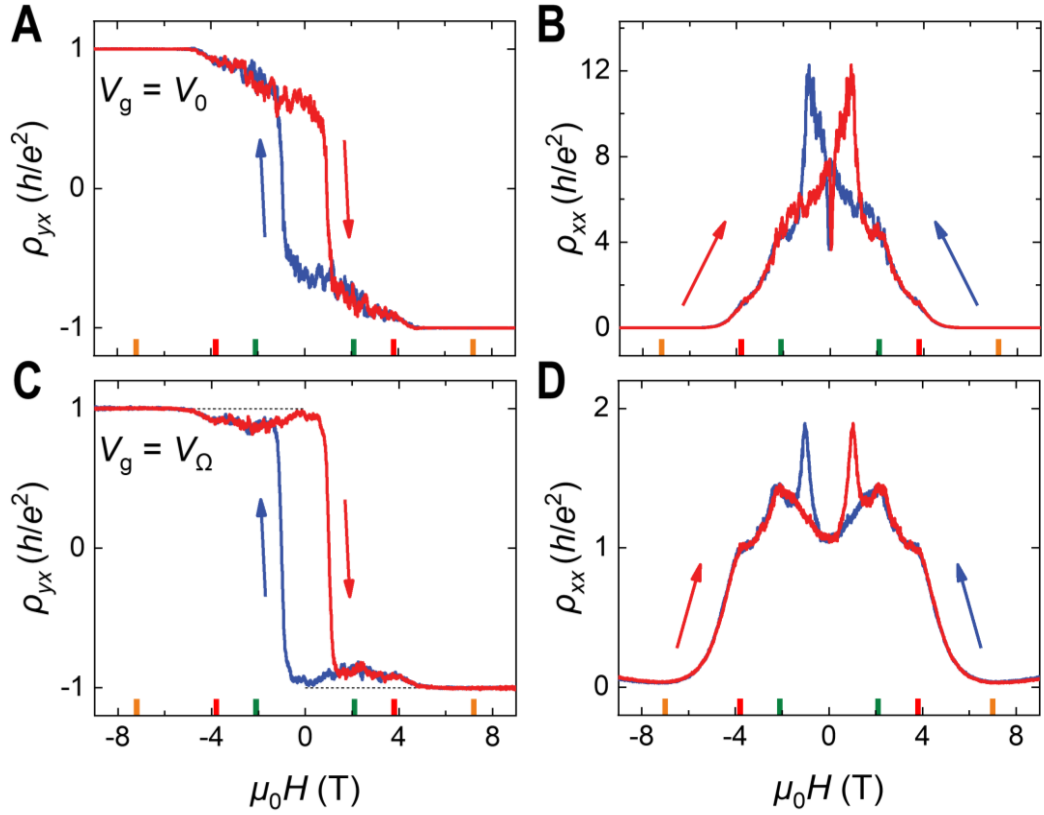

**Fig. 2. QAH and r-QAH effect in a five-layer MnBi$_2$Te$_4$ thin film #1 at $T$ = 20 mK.** (**A** and **B**) Hysteretic $\rho_{yx}$ and $\rho_{xx}$ at $V_0$ as a function of $\mu_0H$. The ZMF $|\rho_{yx}| \approx 0.6$ $h/e^2$ and $\rho_{xx} \approx 7.6$ $h/e^2$ are far from ZMF-QAH effect but common for few-layer MnBi$_2$Te$_4$ device. (**C** and **D**) Hysteretic $\rho_{yx}$ and $\rho_{xx}$ at $V_\Omega$ with respect to $\mu_0H$. Orange, red, and green ticks dividing the curve into four stages, denote ±7.2 T, ±3.8 T, and ±2.1 T, where $|\rho_{yx}| \approx 1.0, 0.9, 0.8$ $h/e^2$, and $\rho_{xx} \approx 0.04, 1.0, 1.3$ $h/e^2$, respectively. The ZMF $\rho_{yx} = 0.9736$ $h/e^2$ and $\rho_{xx} = 1.0756$ $h/e^2$ indicating a r-QAH effect. The data of $\rho_{xx}$ and $\rho_{yx}$ are already symmetrized and antisymmetrized.

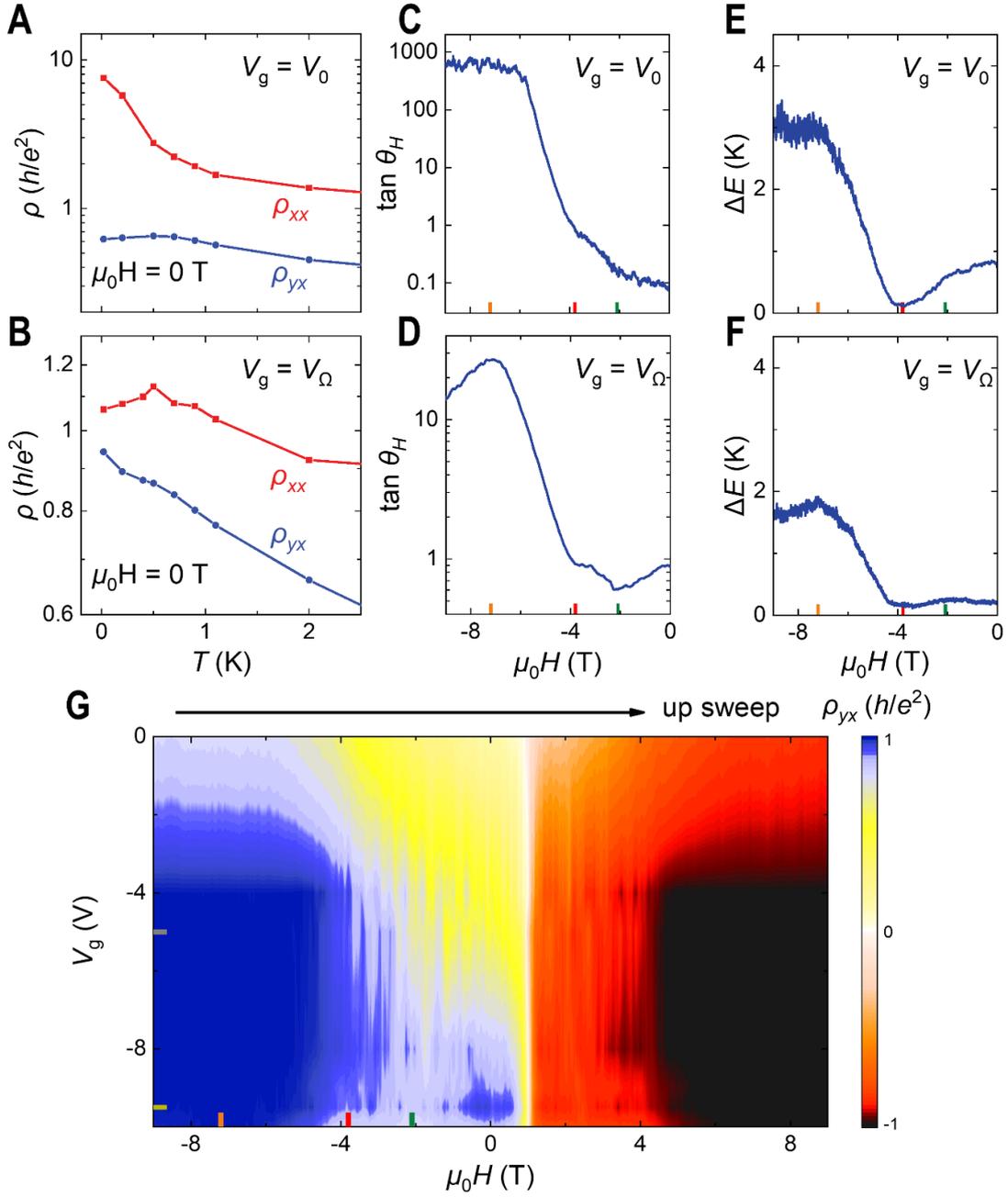

Fig. 3. Comparison of $V_0$ and $V_\Omega$ in the five-layer MnBi$_2$Te$_4$ thin film #1. (A and B) Temperature dependence of $\rho_{xx}$ and $\rho_{yx}$ for the ZMF-insulating state at $V_0$ and the CAI state at 0 T and $V_\Omega$, respectively. (C and D) Tangent of Hall angle $\tan\theta_H = \sigma_{xy}/\sigma_{xx} = \rho_{yx}/\rho_{xx}$ as a function of $\mu_0 H$ for $V_0$ and $V_\Omega$, respectively. Orange, red, and green ticks dividing the curve into four stages, denote -7.2 T, -3.8 T, and -2.1 T. (E and F) Effective energy gap $\Delta E$ as a function of $\mu_0 H$ for $V_0$ and $V_\Omega$, respectively. $\Delta E$ is obtained from fitting the Arrhenius plots. (G) Intensity plot of $\rho_{yx}$ with respect to $\mu_0 H$ and $V_g$ which is collected during $\mu_0 H$ upsweep and subsequently antisymmetrized with downsweep data. Orange, red, and green tick denote -7.2 T, -3.8 T, and -2.1 T. Grey and gold tick denote $V_0$ and $V_\Omega$, respectively. Three QAH regimes is observable: down-HMF-QAH regime ($\mu_0 H$ < -4 T and $V_g$ < -2 V), up-HMF-QAH regime ($\mu_0 H$ > +4 T and $V_g$ < -2 V), and r-QAH regime at 0 T and $V_\Omega$ in the bottom center.

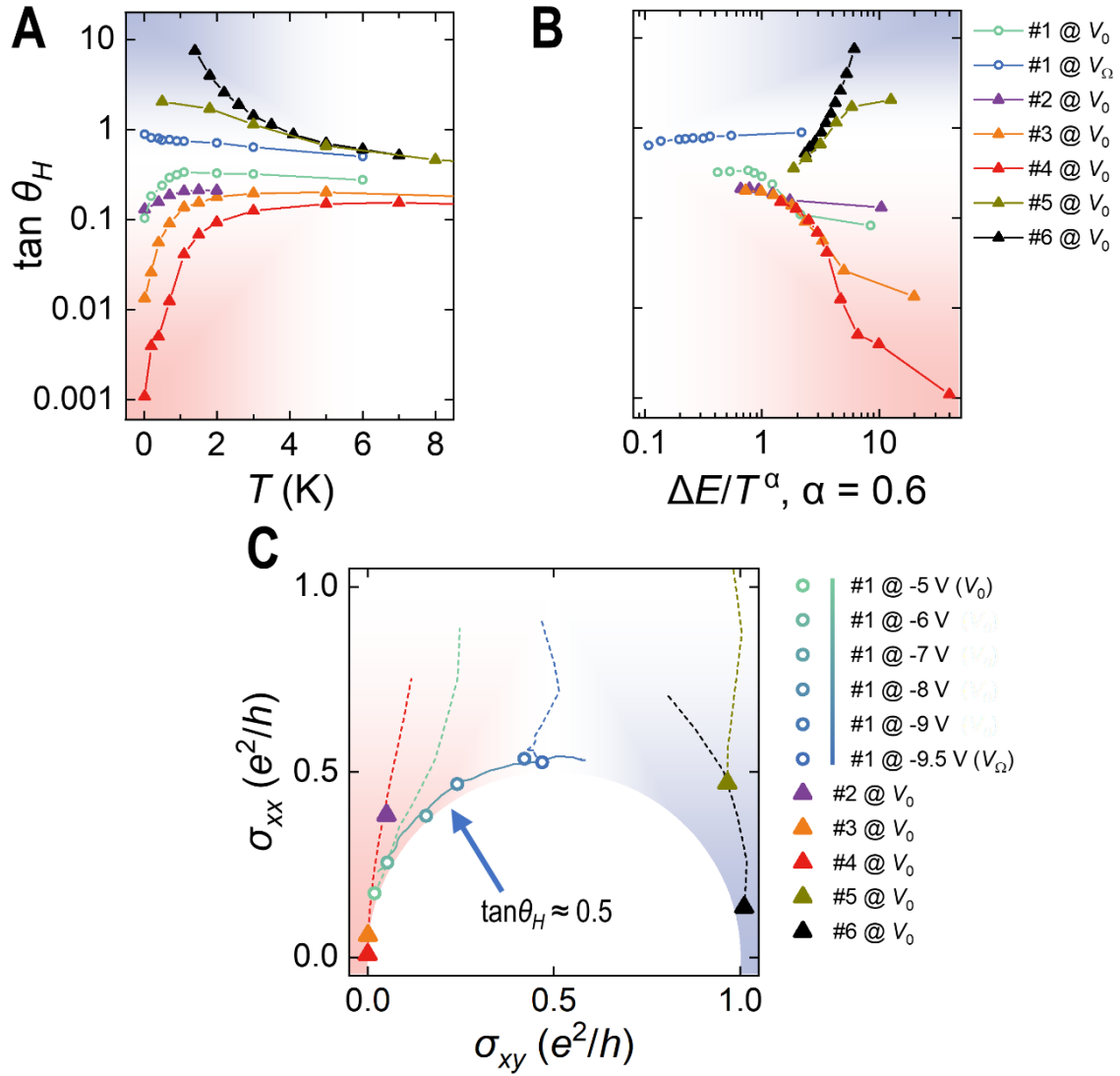

**Fig. 4. Comparison of the five-layer MnBi$_2$Te$_4$ thin film #1 and other samples.** (**A**) Temperature dependence of ZMF tan$\theta_H$ for the thin film #1, additional thin films #2 #3 #4, and exfoliated flakes #5 #6. #5 is the seven-layer flake with ZMF-QAH effect under high pressure [52] and #6 is the five-layer flake with ZMF-QAH effect [6]. (**B**) Scaling analysis of tan$\theta_H$ with respect to $T/\Delta E^\alpha$, where $\Delta E$ is at ZMF and $\alpha = 0.6 \pm 0.05$. At the transition point, the $T$-independent tan$\theta_H \approx 0.5$, obtained from the well-collapsing curve of #2 to #6. (**C**) ZMF parametric plots of ($\sigma_{xy}(\beta)$, $\sigma_{xx}(\beta)$) where $\beta$ evolves $T$ and $V_g$. Dashed lines denote the ZMF-RG flows in ($\sigma_{xy}(\beta)$, $\sigma_{xx}(\beta)$) observed with changing $T$. Solid line denotes ZMF $V_g$ sweep in #1 from $V_0$ to $V_\Omega$. Open circles denote data extracted from $\mu_0 H$ sweep in thin film #1. Triangulars denote thin films #2, #3, #4, and thin flakes #5, #6. Blue arrow indicates the transition point obtained from $T$-independent tan$\theta_H \approx 0.5$. The sky-blue and pink shaded area denote the QAH regime and ZMF-insulating regime, respectively.